\documentstyle[12pt]{article}

\begin{document}
\newcommand{\lartitle}{\ \\  \Large \bf }
\newcommand{\larauthor}{\large \bf }
\newcommand{\laradd}{\normalsize \it }
\newcommand{\BASE}{\baselineskip=15pt}

\begin{titlepage}
\noindent {\large \it Yukawa Institute for Theoretical Physics}
\vspace{-1.2cm}
\begin{flushright}
    YITP-95-2\\
quant-ph/9510001\\
    September 1995
\end{flushright}

\vspace{1cm}
\begin{center}

{\Large \bf         Squeezed States for General Multiphoton Systems:}\\
{\Large \bf         Towards the Displacement-operator Formalism  }

\vspace{1cm}

{\large \bf          Hong-Chen Fu\footnote{ JSPS Fellow. On leave of
                              absence from
                              Institute of Theoretical Physics,
                              Northeast Normal University,
                              Changchun130024, P.R.China.
                              E-mail: hcfu@yukawa.kyoto-u.ac.jp}
                    \ and Ryu Sasaki\footnote{
                              Supported partially by the grant-in-aid for
                              Scientific Research, Priority Area 231
                              ``Infinite Analysis'' and General Research (C) in
                              Physics, Japan Ministry of Education.} }\\
\vspace{0.5cm}

{\it                Yukawa Institute for Theoretical Physics, Kyoto
                    University}\\
{\it                Kyoto 606-01, Japan}
\end{center}
\vspace{2cm}

\begin{abstract}
We propose a displacement-operator approach to some aspects of squeezed
states for general multiphoton systems.  The explicit
displacement-operators of the squeezed vacuum and the coherent states
are achieved and expresses  as the ordinary exponential
form. As a byproduct the coherent states of the
$q$-oscillator are obtained by the {\it usual exponential}
displacement-operator.
keywords:  squeezed states,  coherenr states, squeezed vacuum,
general multiphoton systems, displacement-operator, $q$-oscillator\\

\end{abstract}
\end{titlepage}

%************************** Text Begins here ******************************

\newcommand{\ad}{A^{\dagger}}
\newcommand{\at}{A^{\dagger 2}}
\newcommand{\be}{\begin{equation}}
\newcommand{\ee}{\end{equation}}
\newcommand{\NN}{{\cal N}}
\newcommand{\AAA}{{\cal A}^{\dag}}
\newcommand{\bn}{\begin{eqnarray}}
\newcommand{\en}{\end{eqnarray}}
\newcommand{\no}{\nonumber}
\newcommand{\ff}{&\ &\hspace{-0.8cm}}
\newcommand{\rr}{\rangle}
\newcommand{\rref}[1]{(\ref{#1})}
\newcommand{\ts}{\thinspace}
\newcommand{\CC}{\widehat{C}}
\newcommand{\DD}{\widehat{D}}
\def\twb{|\!|}
\newcommand\p{^\prime}
\newcommand\pp{^{\prime\prime}}
%-----------------------------------------

\section{Introduction}

It is well known that there are three equivalent definitions of the
coherent states of the harmonic oscillators, that is, (1) the
displacement-operator acting on the vacuum states, (2) the eigenstates of
the
annihilation operator and (3) the minimum uncertainty states.
Generalizations to the coherent states of  arbitrary Lie groups were
extensively studied in the literature \cite{book,feng}. However,
their extension to the squeezed states \cite{yuen}, which became
more and more interesting in  quantum optics \cite{work} and
gravitational
wave detection\cite{cave}, gave the equivalent results
only for the harmonic oscillator. The minimum uncertainty method works
well for both the coherent and squeezed states for any symmetry
systems \cite{nie3,nie4} and the ladder-operator squeezed states
for general systems
are described in \cite{nie1}. Both methods are closely connected as
discussed in \cite{nie1}. Now the exception is that there is no
general approach to the displacement-operator squeezed states for
the general systems although there are some works toward this goal
\cite{nie2}.

In this letter we shall investigate squeezed states for the general
multiphoton multimode systems (see Eqs.(\ref{aacom}),\ts (\ref{re})
and (47)).
We first formulate the algebras of multiphoton creation and
annihilation operators, and then discuss the squeezed states from
the ladder-operator definition. For the
zero squeezing case, we derive the explicit coherent state, which is
written as an ordinary exponential displacement-operator acting on the
vacuum state. The squeezed vacuum of this system is also explicitly
obtained as an exponential displacement-operator, which is the {\it
squeeze operator}, acting on the vacuum state. However, the product of
coherent and the squeeze operators does not bring us the squeezed states
equivalent to those obtained by the ladder-operator method.
As a byproduct, the coherent states of the
$q$-deformed oscillator \cite{hcfu} are obtained and expressed as the
usual exponential displacement-operator acting on the vacuum states.

\section{Multiphoton algebra}

We first recall the ladder-operator approach of Nieto
and Truax. They describe that {\it the general ladder-operator squeezed
states are the eigenstates of a linear combination of the lowering
and raising operators} \cite{nie1}
\be
      \left(\mu A+ \nu \ad \right)|\beta\rr=\beta |\beta\rr, \label{eigeq}
\ee
where  $\mu$ and $\nu$ are  complex constants,
$A$ and its hermitian conjugate $\ad$ are the
lowering (annihilation) and raising (creation) operators,
respectively. They satisfy the commutation relation
\be
      \left[A, \ad \right]=C, \label{aacom}
\ee
in which $C$ is an operator to be specified later.
We consider the eigenvalue equation \rref{eigeq}\ts in the parameter region
$|z|=|-\nu/\mu|<1$.
This means that the operator $\mu A+ \nu \ad$ has more annihilation
than creation  operators (see \rref{vv}).
Here we consider only the single mode case and generalization to
multimode case will be discussed in section 4.
In quantum optics  $A$ is usually the multiphoton lowering operator
in the form
\be
       A= f(N)a^m,
\ee
where $a$ and $a^{\dagger}$ are the
annihilation and creation
operators of the photon satisfying $\left[a,
a^{\dagger}\right]=1$, $N=a^{\dagger}a$,  and $m$ is a positive integer.
As usual the Fock states of the oscillator $a$ and $a^{\dagger}$ are
denoted by $|n\rr$, $n=0,1,\ldots$, $a|0\rr=0$, $a|n\rr=\sqrt{n}|n-1\rr$,
$a^{\dagger}|n\rr=\sqrt{n+1}|n+1\rr$.
The function $f(N)$ specifies the intensity dependent coupling,
which is in general complex and we assume that $f(x)$ does not have zeros
at non-negative integer values of $x$.
By using $aa^{\dagger}=N+1$, $a^2(a^\dagger)^2=(N+1)(N+2)$, etc, we obtain
\bn
    A \ad &=& (N+1)(N+2)\cdots (N+m) f(N)f^*(N), \label{aada} \\
    \ad A &=& (N-m+1)(N-m+2)\cdots N f(N-m) f^*(N-m).\label{aaad}
\en
It is obvious that we only need to restrict our discussions to the sector
$S_i$ ($i=0,1,\cdots,m-1$) spanned by the Fock states $|nm+i\rr$ ($n$
non-negative integers). Introducing the
{\it multiphoton number operator} $\NN_i \equiv \NN$
($i=0,1,\cdots,m-1$) in the sector $S_i$
\be
      \NN = \frac{1}{m}\left(N-i\right),\ \ \ i=0,1,\cdots,m-1,
\ee
and $F(\NN+1)\equiv (m\NN+1+i)\cdots(m\NN+m+i)f(m\NN+i)f^*(m\NN+i)$,
we can recast the
system \rref{aacom},\ts \rref{aada},\ts \rref{aaad}\ts in the following
form
\bn
\ff      A\ad = F(\NN+1),\ \ \  \ad A=F(\NN),\nonumber \\
\ff       \left[\NN, \ad\right]=\ad,\ \ \
         \left[\NN, A\right]=-A.   \label{re}
\en
These are the starting relations of this letter, which we call
{\em general intensity dependent $m$-photon algebra}.
\footnote{A similar algebra appeared in Heisenberg's theory of
non-linear
spinor dynamics \cite{RSW}, Eq.(9.6). We thank D. Fairlie for calling our
attention
to this paper.}
With  these definitions, the operator $C$ in \rref{aacom}\ts is
$C=F(\NN+1)-F(\NN)$.
Note that the r.h.s.
of \rref{aaad} vanishes on the Fock states $|n\rr$ for $0\leq n \leq m-1$,
which implies $F(0)=0$ in each sector.

The system \rref{re} is general enough to cover many interesting examples:
The case  $m=1$ and $f(N)=1$ is the harmonic oscillator;
for $m=1$ and $f(N)=\sqrt{2k+N}$ ($k\ge0$) we recover the
Holstein-Primakoff
realization of the $su(1,1)$ algebra \cite{hp} by identifying
\be
    K^+ \equiv \ad,\ \ \ \   K^-\equiv A,\ \ \ \  K^0 \equiv k+N.\label{hp}
\ee
The square realization of $su(1,1)$ corresponds to: $m=2$ and $f(N)=1/2$
\bn
    K^+    & \equiv & \frac{1}{2}(a^{\dagger})^2,\ \ \ \
    K^-     \equiv \frac{1}{2} a^2,\nonumber \\
    K^0    & \equiv &\frac{1}{2}\left(N+\frac{1}{2}\right)
            =      \left\{ \begin{array}{ll}
                                \NN_0 + \frac{1}{4} & \mbox{in}\ S_0,\\
                                \NN_1 + \frac{3}{4} & \mbox{in}\ S_1.
                           \end{array}
                   \right. \label{squ}
\en
Both realizations \rref{hp} and \rref{squ} satisfy the $su(1,1)$ defining
relations
\bn
   \left[K^+, K^-\right]&=&-2K^0,\ \ \
   \left[K^0, K^+\right]=K^+  ,\nonumber  \\
   \left[K^0, K^-\right]&=&-K^-.
\en

It is convenient to introduce an orthonormal basis for $S_i$
\be
     \twb n\rr=\frac{1}{\sqrt{F(n)!}}\left(\ad\right)^n \twb 0\rr,
\ee
where $\twb 0\rr=|i\rr$ is the vacuum state of sector $S_i$ satisfying
$A\twb 0\rr=\NN\twb 0\rr=0$ and
$ F(n)!\equiv F(n)F(n-1)\cdots F(1),\ \ \  F(0)!\equiv 1 $.
On this basis we have
\bn
    \ad \twb n\rr &=& \sqrt{F(n+1)}\twb n+1\rr,\no \\
    A   \twb n\rr &=& \sqrt{F(n)}\twb n-1\rr,\no \\
    \NN \twb n\rr &=& n \twb n\rr.  \label{reps}
\en

It is very tempting to apply the idea of  the system \rref{re} to the
multiphoton ($m$-photon) coherent states of the $q$-deformed
oscillator. Let us choose  ($q$ is a {\em real} deformation parameter)
\be
   f(N)\equiv \left\{ \frac{1}{(N+1)\cdots (N+m)}\left[
              \frac{N}{m}+1\right]\right\}^{\frac{1}{2}}\label{fqdef},
\ee
where $[x]\equiv (q^x-q^{-x})/(q-q^{-1})$,
and define
\bn
\ff      b_q \equiv A=f(N)a^m,\ \ \
         b_q^{\dagger} \equiv \ad=(a^{\dagger})^mf(N), \nonumber \\
\ff      N_q        \equiv \NN+\frac{i}{m}. \label{q}\ \ \
\en
Then by using \rref{aada}\ts and \rref{aaad}\ts we would obtain {\em
formally}
the following  relations
\bn
   b_qb_q^{\dagger}  &=&  {(N+1)(N+2)\cdots (N+m)\over{(N+1)(N+2)\cdots
                       (N+m)}}[N_q+1]\nonumber \\
                  &=&  [N_q+1], \label{BBdag} \\
   b_q^{\dagger}b_q  &=&  {(N-m+1)(N-m+2)\cdots N\over{(N-m+1)(N-m+2)\cdots
                       N}}[N_q]\nonumber  \\
                  &=&  [N_q], \label{bdagb}  \\
\   [N_q,& b_q^{\dagger}&]= b_q^{\dagger}, \quad \ [N_q, b_q]=-b_q. \label{Nbb}
\en
Eqs.\rref{BBdag},\rref{bdagb},\rref{Nbb} are in fact a multiphoton
realization of the $q$-oscillator,
\be
   b_qb_q^{\dagger}-qb_q^{\dagger}b_q=q^{-N_q}. \label{qboson}
\ee
(The case $m=1$ was found in \cite{song}. See also \cite{KDa}, \cite{CZ}
and \cite{MPa}.) It should be remarked that the
eigenvalues of $N_q$ are not integers except for the $i=0$ sector.

By close inspection, however, one finds that the relation \rref{bdagb}
$$
   b_q^{\dagger}b_q = [N_q]
$$
is not true on the vacuum in each sector $S_i$ ($i\ge1$ and
$m>1$). Obviously the vacuum of the $i$-th sector $\twb 0\rr=|i\rr$
vanishes
when applied by $b_q$,
\be
   b_q\twb 0\rr=f(N)a^m|i\rr=0, \quad i=0,1,\ldots,m-1. \label{bkill}
\ee
On the other hand, as remarked above, $[N_q]\twb 0\rr=[{i\over m}]\twb 0\rr$ is
{\it
non-vanishing\/} for $i\ge1$. This apparent inconsistency is caused by
$0/0=1$ in \rref{bdagb}, since $N-i$ in the numerator and denominator
vanish
on $\twb 0\rr=|i\rr$. To sum up, the relations
\rref{qboson}\ts and \rref{bdagb}\ts are broken only by the vacuum
expectation value and all the other relations are correct.
It would be very interesting if one could find physical applications
of the ``spontaneously broken'' {\it $q$-deformed multi-photon coherent
states}.
\footnote{If we introduce the intensity and sector-dependent multiphoton
coupling then we can obtain the $q$-deformed oscillator in each sector.
Namely, if we define
$a_q=\sqrt{[\NN+1]\over{(N+1)\cdots(N+m)}}\ts a^m$ in each sector, then it
is
easy to see that
$a_qa_q^{\dagger}=[\NN+1]$ and $a_q^{\dagger} a_q=[\NN]$ are satisfied as
operator
equations.}

\section{Squeezed states}

Now we shall solve the eigenvalue equation (1). We expand the state
$|\beta\rr$ as
\be
    |\beta\rr = \sum_{n=0}^{\infty}C_n \twb n\rr,
\ee
Then equation (1) leads to the following recursion relations
\bn
\ff     C_{n+1}=z\sqrt{\frac{F(n)}{F(n+1)}}C_{n-1}+
                \frac{\beta}{\mu\sqrt{F(n+1)}}C_n, \nonumber \\
\ff     C_1=\frac{\beta}{\mu \sqrt{F(1)}}C_0,  \label{recur}
\en
where $z=-\nu/\mu$. We have not been able to obtain a closed expression
of $C_n$ for the general case. We now consider some special cases.

\subsection{Squeezed vacuum and squeeze operator}

Let us first consider the squeezed vacuum $|v\rr$ annihilated by $\mu A
+\nu \ad$
\be
     \left(\mu A +\nu \ad \right)|v\rr = 0,  \label{bv}
\ee
and express it in terms of an exponential displacement-operator
(squeeze operator) acting on the vacuum state.
In this case with $\beta=0$, the $C_n$'s are easily obtained as
\be
     C_{2k+1}=0,\ \ \  C_{2k}=C_0 z^k
                       \sqrt{\frac{F(2k-1)!!}{F(2k)!!}},
\ee
where $F(2k)!!=F(2k)F(2k-2)\cdots F(2)$,
$F(2k-1)!!= F(2k-1)F(2k-3)\cdots F(1)$ and $F(0)!!=F(-1)!!\equiv 1$.
Then we have
\bn
     |v\rr &=& C_0\sum_{k=0}^{\infty}z^k
               \sqrt{\frac{F(2k-1)!!}{F(2k)!!}}\twb 2k\rr   \nonumber \\
           &=& C_0 \sum_{k=0}^{\infty}z^k
               \frac{\left(\at\right)^k}{F(2k)!!}\twb 0\rr.  \label{vv}
\en
It is easy to check that the above infinite series converges if $|z|<1$
under mild assumptions on the asymptotic behavior of $f(x)$, e.g.,
$f(x)\simeq x^\alpha$ for $x\to\infty$.
By making use of the following identity
\be
     \left( \frac{\NN}{F(\NN)}\at\right)^k = \left(\at\right)^k
     \frac{\NN+2}{F(\NN+2)}\cdots \frac{\NN+2k}{F(\NN+2k)},
\ee
we can rewrite \rref{vv} as
\bn
     |v\rr &=& C_0
              \sum_{k=0}^{\infty}\frac{1}{k!}
              \left(\frac{z}{2}\at\right)^k
              \frac{(\NN+2)\cdots (\NN+2k)}
              {F(\NN+2)\cdots F(\NN+2k)} \twb 0\rr   \no  \\
          &=& C_0 \exp{\left(\frac{z \NN}{2F(\NN)}\at\right)}\twb 0\rr.
                        \label{vsol}
\en
Following the terminology of the oscillator, the operator
\be
     S(z)=C_0 \exp{\left(\frac{z\NN}{2F(\NN)}\at \right)}
\ee
is referred to as the {\it generalized} squeeze operator.

\subsection{Multiphoton coherent states}

Next let us consider the case $\nu=z=0$. In this case the equation (1)
reduces to the eigenvalue equation of the annihilation operator $A$ and the
resulting states are the coherent states. From Eq.\rref{recur} with
$z=0$ we can easily obtain the $C_n$'s as
\be
        C_n=\frac{C_0\alpha^n}{\sqrt{F(n)!}},
\ee
where $\alpha=\beta/\mu$. Then the eigenstate $|\beta/\mu\rr\equiv
|\alpha\rr$ is obtained as
\be
       |\alpha\rr = C_0 \sum_{n=0}^{\infty} \frac{\alpha^n}{F(n)!}
                    (\ad)^n \twb 0\rr.\label{coherent}
\ee
In terms of the so-called deformed exponential function
$\exp_F (x)=\sum_{n=0}^{\infty}\frac{x^n}{F(n)!}$ one can express the
coherent state as
$|\alpha\rr=C_0\exp_F(\alpha\ad)\twb 0\rr$.
Here we   would rather like to use the usual exponential
displacement-operator.
To do this, we use the following relation
\be
    \left( \frac{\NN}{F(\NN)} \ad \right)^n =
    (\ad)^n \frac{\NN+1}{F(\NN+1)} \cdots  \frac{\NN+n}{F(\NN+n)},
\ee
and rewrite the Eq.\rref{coherent} in the following form
\bn
    |\alpha\rr &=& C_0 \sum_{n=0}^{\infty} \frac{\alpha^n}{n!}
                   \left(\frac{\NN}{F(\NN)}\ad \right)^n \twb 0\rr \no \\
               &=& C_0 \exp\left(\frac{\alpha\NN}{F(\NN)}\ad \right)
                   \twb 0\rr.
                   \label{cs}
\en
The operator
\be
   D(\alpha)\equiv C_0 \exp\left(\frac{\alpha\NN}{F(\NN)}\ad \right)
\ee
is the coherent displacement-operator in the form of the ordinary
exponential function.

Let us remark that two displacement operators $\exp_F(\alpha\ad)$ and
$\exp\left(\frac{\alpha\NN}{F(\NN)}\ad\right)$ are
essentially different, although they give rise to the same coherent
states by acting on the vacuum state.

We would like to mention that the coherent state \rref{cs} is
the same as that given in [10]. Indeed, one
can easily show that
\be
   \left[A, \frac{\NN}{F(\NN)}\ad \right]=1
\ee
and therefore
\be
     [A,\ D(\alpha)]=\alpha D(\alpha)   \label{bdrel},
\ee
which leads to Eq.\rref{cs} immediately.

The next example is the square realization of  $su(1,1)$, \rref{squ}.
In this case
$(\NN/F(\NN))\ad =\frac{1}{N-1}(a^{\dagger})^2$ in the sector $S_0$ and
$(\NN/F(\NN))\ad=\frac{1}{N}(a^{\dagger})^2$ in the sector $S_1$. Then
we obtain the well-known two-component coherent states of
$su(1,1)$:
\bn
  |\alpha\p\rr_0&=& C_0
                   \sum_{k=0}^{\infty}\frac{{\alpha\p}^{2k}}
                   {\sqrt{(2k)!}}|2k\rr, \quad \quad \quad \quad \
                   \alpha\p=\sqrt{2\alpha}, \no  \\
  |\alpha\p\rr_1&=& C_0\p
                   \sum_{k=0}^{\infty}\frac{{\alpha\p}^{2k+1}}
                   {\sqrt{(2k+1)!}} |2k+1\rr, \quad
                   C_0\p=C_0{\alpha\p}^{-1}.
\label{jjj}
\en
Note that in (\ref{jjj})
we use the usual Fock states.

It should be mentioned that the coherent states of the $q$-deformed
oscillators ($m=1$ case) can also be obtained by inserting \rref{fqdef} and
\rref{q}  into the
general formula \rref{cs}. We have advanced the understanding of the
problem on two points, that is, (i)  the displacement-operator is
expressed by the usual exponential form \cite{KDa},
not the so-called $q$-deformed
exponential function \cite{hcfu,KDa}; (ii) the $q$-deformed oscillator admits
the
multi-component squeezed and coherent states through its multiphoton
realization \rref{q}\ts but the relationship is broken by the vacuum
expectation value. The above presentation  can also be applied to the
case of the quantum algebra $su(1,1)_q$.

\subsection{Squeezed states}

We have constructed the coherent displacement-operator $D(\alpha)$
and the squeeze operator $S(z)$. For the oscillator, we know that the
state $D(\alpha)S(z)|0\rr$
\bn
      |\alpha,z\rr &\equiv D(\alpha)S(z)|0\rr&=
                  C_0 e^{\alpha a^{\dagger}}
                    e^{{z\over2} a^{\dagger 2}}|0\rr  \no \\
                  &\stackrel{\mbox{\footnotesize normalization}}
                    {=\!=\!=\!=\!=\!=\!=\!=\!=}&
                    e^{\alpha a^{\dagger}-\alpha^* a}
                    e^{{z\over2} a^{\dagger 2}-{z^*\over2} a^2}|0\rr
\en
is just the squeezed state, which is also
an eigenstate of $\mu a + \nu a^{\dagger}$. However, for the general
case here, the state $D(\alpha)S(z)\twb 0\rr$ are not the squeezed state
equivalent to the ladder-operator definition, namely, it is
not an eigenstate of $\mu A +\nu \ad$. The coherent
displacement-operator $D(\alpha)$ is a {\it good} operator in the
sense that it enjoys the following property
\be
      D(-\alpha)AD(\alpha)=A+\alpha,
\ee
which can be easily obtained from the following identity
\be
      e^{\xi P}Qe^{-\xi P}=Q+\xi[P,Q]+\frac{\xi^2}{2!}[P,[P,Q]]+\cdots .
\ee
However, the squeeze operator $S(z)$ does not keep the
Holstein-Primakoff/Bogoliubov
transformation, namely
\be
     S^{-1}(z)AS(z)\neq \mu A + \nu \ad.
\ee
This is why the state $D(\alpha)S(z)\twb 0\rr$ is not an eigenstate of
$\mu A+\nu \ad$, as argued in the paper \cite{nie2}.

Now let us consider another special case, $\nu=-\beta^2$ or
$z=\mu^{-1}\beta^2$.
In this case it is easy to see that the coefficient $C_n$ can be written as
$C_n = \CC(n,\mu) \beta^n$ and the $\CC(n,\mu)$ satisfying the following
relation
\bn
     \CC(n+1,\mu)&=&\frac{1}{\mu}\sqrt{\frac{F(n)}{F(n+1)}}\CC(n-1,\mu)
     +\frac{1}{\mu\sqrt{F(n+1)}}\CC(n,\mu), \no \\
     \CC(1,\mu)&=&\left(\mu\sqrt{F(1)}\right)^{-1}\CC(0,\mu),
\en
where $\CC(0,\mu)\equiv C_0$ is a constant independent of $\mu$. Now we
introduce a new symbol by
\be
   \DD(n,\mu)\equiv \CC(n,\mu)\sqrt{F(n)}!,
\ee
then $\DD(n,\mu)$ is determined by
\bn
\ff   \DD(n+1,\mu)=\mu^{-1}\left(\DD(n,\mu)+F(n)\DD(n-1,\mu)\right),\no \\
\ff   \DD(1,\mu)=\mu^{-1}C_0.
\en
The squeezed state $|\beta,\mu\rr$ is obtained as
\bn
   |\beta,\mu\rr &=& \sum_{n=0}^{\infty}\DD(n,\mu)\frac{\beta^n}{\sqrt{F(n)!}}
                 \twb n\rr  \no \\
             &=& \sum_{n=0}^{\infty}\DD(n,\mu)\frac{\beta^n}{F(n)!}
                 ( \ad)^n \twb 0\rr   \no \\
             &=& \sum_{n=0}^{\infty}\frac{1}{n!}\left(\beta \frac{\NN}
                 {F(\NN)}\ad\right)^n \DD(\NN+n,\mu)\twb 0\rr   \no \\
             &=& \sum_{n=0}^{\infty}\frac{1}{n!}\DD(\NN,\mu)
                 \left(\beta \frac{\NN}
                 {F(\NN)}\ad\right)^n \twb 0\rr   \no \\
             &=& \DD(\NN,\mu)\exp\left(\beta\frac{\NN}{F(\NN)}
                 \ad\right) \twb 0\rr \no \\
             &=& \DD(\NN,\mu)D(\beta)\twb 0\rr.
\en
It is easy to show that the above infinite series converges for $|1/\mu|<1$
and $|\beta|<1$
under about the same assumptions on the asymptotic behavior of $f(x)$ as
before.
It is interesting to note for $F(n)=n^2$ and $\mu=1$, the recursion
relations can be solved explicitly: $\DD(n,1)/C_0=n!$. In this case the
creation part of the operator $\mu A+\nu\ad$ ($\nu=-\beta^2$)
can be considered as a small perturbation to the annihilation part.

Recall that for the oscillator, the squeezed state can also be written
as
\be
     |\alpha, z\rr=D(\alpha)S(z)|0\rr \equiv S(\gamma)D(\alpha)|0\rr,
\ee
where $\gamma=\alpha\cosh r-{\alpha^*}e^{i\theta}\sinh r$
($z=re^{i\theta}$).
The operator $\DD(\NN,\mu)$ seems to play the role of a squeeze operator,
but not exactly. It contains only
the number operator $\NN$. In comparison with the oscillator case, one
finds that $\DD(\NN,\mu)$ is obviously different from  the squeeze
operator $S(z)$.

Although the states $D(\alpha)S(z)\twb 0\rr$ and $S(z)D(\alpha)\twb 0\rr$
are not the eigenstates of $\mu A+\nu \ad$, they might be important
quantum states in quantum optics. Investigation of their non-classical
properties will be of significance. We shall consider this problem
in a separate paper.

\section{Multimode generalization}

The above formalism can be easily generalized to the multimode case. For
simplicity we consider only the two-mode case \cite{PA}. Generalization to
multimode is straightforward.
Consider the two-mode photon field described by two independent modes
\be
    [a, a^{\dagger}]=1,\ \ [b, b^{\dagger}]=1,
\ee
and introduce a two-mode multiphoton annihilation operator
\be
    A=f(N_1,N_2)a^m b^n,
\ee
where $N_1=a^{\dagger}a,\ N_2=b^{\dagger}b$, $f$ is an arbitrary
function with $f(n_1,n_2)\neq 0$ for $ n_1,n_2$ non-negative
integers. Note that $f$ is not necessarily written as $f(N_1,N_2)
=f_1(N_1)f_2(N_2)$. Then we have
\bn
\ff    A\ad=F(\NN_1+1,\NN_2+1), \ \ \
       \ad A= F(\NN_1,\NN_2), \no \\
\ff    [\NN_i, \ad]=\ad,\ \ \ [\NN_i, A]=-A, \ \ \ (i=1,2),
\en
where
\bn
    \NN_1&\equiv& \frac{1}{m}(N_1-i), \ \
       \NN_2\equiv \frac{1}{n}(N_2-j), \ \
       (0\leq i\leq m-1,\ 0\leq j\leq n-1),  \no \\
    F(\NN_1+1,\NN_2+1) &\equiv&
           (N_1+1)\cdots(N_1+m)(N_2+1)\cdots(N_2+n) f(N_1,N_2)f^*(N_1,N_2)\no
\\
    &\equiv& (m\NN_1+i+1)\cdots (m\NN_1+i+m)(n\NN_2+j+1)\no \\
    &&\cdots(n\NN_2+j+n) f(m\NN_1+i,n\NN_2+j)f^*(m\NN_1+i,n\NN_2+j).
\en
This algebra is defined on a subspace $\bar{S}_{ij}$ of the
sector $S_{ij}$. A convenient orthonormal basis of $\bar{S}_{ij}$ is given
 by ($ k=0,1,2,\cdots $)
\be
       \twb k\rr \equiv \frac{1}{\sqrt{F(k,k)!}}
                    (\ad)^k |i,j\rr \propto |km+i,kn+j\rr.
\ee
The representation on $\bar{S}_{ij}$ is
\bn
     \ad \twb k\rr &=& \sqrt{F(k+1,k+1)}\twb k+1\rr, \no \\
     A   \twb k\rr &=& \sqrt{F(k,k)}\twb k-1\rr, \no \\
   \NN_1 \twb k\rr &=& \NN_2 \twb k\rr = k \twb k\rr.   \label{rree}
\en

We consider the eigenvalue equation
\be
     \left(\mu A+\nu \ad\right)|\beta\rr=\beta|\beta\rr.
\ee
These states are degenerate. The degeneracy can be lifted by assuming
that the $(m+n)$ photons are either created or annihilated
together. This means the following conservation law
\be
     \left(\NN_1 - \NN_2\right)|\beta\rr = 0.
\ee
In the representation \rref{rree} the condition is fulfilled
automatically.

By identifying $F(k,k)$ here with $F(k)$ in section 2, the representation
\rref{rree} takes the same form as \rref{reps}. So, formally, the squeezed
states can be investigated in the same manner as those in section 2.

\section{Conclusion}

So far we have described a displacement-operator formalism of the
squeezed states for the general symmetry system. The coherent
displacement-operator and the squeeze operator are explicitly
constructed. Although these two operators are equivalent to the
ladder-operator definition, their product does not bring us the
squeezed states consistent with the ladder-operator definition, due to
the lack of Holstein-Primakoff/Bogoliubov transformation.
However, we can expect that the
states $D(\alpha)S(z)\twb 0\rr$ and $S(z)D(\alpha)\twb 0\rr$ play
important roles in quantum optics and it is a good challenge to study
their non-classical properties.

As a special case of this formalism,
we obtain the coherent displacement-operator of the
$q$-deformed oscillator, which takes the form of the {\it  usual exponential
function} form.

This formalism can also be applied to the systems with self-similar
potentials \cite{self}. In these systems, the dynamical symmetry
algebra (called $q$-ladder algebra) belongs to  the general category
\rref{re}.
In the isospectral  oscillator hamiltonian systems \cite{MFHN}, the creation
and
the annihilation operators also enjoy the algebraic structure \rref{re}
and therefore their coherent states can be studied as a special case
of this letter. In particular, by making use of the techniques in
section 3, some interesting results on the relation of this
coherent and squeezed
states with those of the oscillator can be found. The concrete properties of
the various coherent states arising from them
will be published in a separate paper.

%%%%%%%%%%%%%%%%%%%%%%%%%%%%%%%%%%%%%%%%%%%%%%%%%%%%%%%%%%%%%%%%%%%%%%%%%

\section*{Acknowledgments}

We thank D.\ts Fairlie, P.\ts Kulish, M.\ts Pillin and C.\ts Zachos
for useful comments. H.\ts C.\ts Fu is grateful to Japan Society for
Promotion of Science (JSPS) for the fellowship.
He is also supported in part by the National Science Foundation of
China.

%%%%%%%%%%%%%%%%%%%%%%%%%%%%%%%%%%%%%%%%%%%%%%%%%%%%%%%%%%%%%%%%%%%%%%%%%%%%%%

\def\JP{{J.\ts Phys.\ts}}
\def\JMP{{J.\ts Math.\ts  Phys.\ts}}
\def\PL{{Phys.\ts Lett.\ts}}
\def\PR{{Phys.\ts Rev.\ts}}
\def\PRD{{Phys.\ts Rev.\ts D\ts}}
\def\PRL{{Phys.\ts Rev.\ts Lett.\ts}}
\def\RMP{{Rev.\ts Mod.\ts Phys.\ts}}

%%%%%%%%%%%%%%%%%%%%%%%%%%%%%%%%%%%%%%%%%%%%%%%%%%%%%%%%%%%%%%%%%%%%%%%%%%%%%%

\end{document}